\documentclass{aa}
\usepackage{graphicx}
\usepackage{txfonts}
\usepackage{color}
\begin{document}
   \title{NACO Polarimetric Differential Imaging of TW Hya:}

   \subtitle{A Sharp Look at the Closest T Tauri Disk\thanks{
   The data presented here were obtained during the commissioning of the 
   NACO instrument at the Paranal Observatory of ESO.}
   }

   \author{D. Apai\inst{1}
          \and
          I. Pascucci\inst{1}
	  \and
	  W. Brandner\inst{1}
	  \and
	  Th. Henning\inst{1}
	  \and
	  R. Lenzen\inst{1}
	  \and
	  D. E. Potter\inst{2}
	  \and
	  A.-M. Lagrange\inst{3}
	  \and
	  G. Rousset\inst{4}
	  }

   \offprints{D. Apai, \email{apai@mpia.de}}

   \institute{Max Planck Institute for Astronomy, K\"onigstuhl 17, Heidelberg, D-69117 Germany
                   \and
	      Steward Observatory, University of Arizona, 933 N. Cherry Avenue, Tucson, AZ 85721
             		\and
	     Laboratoire d'Astrophysique, Observatoire de Grenoble, 414, rue de la piscine, BP 53, 38041 Grenoble Cedex 9, France
	     \and
	     ONERA, BP 72, 29 avenue de la Division Leclerc, 92322 Ch\^atillon Cedex, France
	     }

   \date{Received 11 Sep, 2003; accepted 3 Nov, 2003}

   \abstract{We present high-contrast imaging data on the disk of the classical T Tauri star TW Hya.
    The images were obtained through the polarimetric differential imaging technique 
with the adaptive optics system NACO. Our commissioning data show the 
presence of polarized disk emission between 0.1\arcsec{} and 1.4\arcsec{} from the star. 
We derive the first Ks-band radial polarized intensity distribution. We show that the polarized
intensity compares well to shorter wavelengths surface brightness observations and confirm 
the previously reported gradual slope change around 0.8\arcsec{}.
These results show the potential of the new polarimetric differential imaging technique
at 8m-class telescopes to map the inner regions of protoplanetary disks.
   }

   \maketitle
%

\section{Introduction}

The simulations of viscous disks  with an embedded Jupiter-mass
 planet (see, e.g. D'Angelo et al. \cite{dan03}) predict dramatic changes in the disk structure: gaps are being opened
and shock waves follow the path of the orbiting body. 
 Detecting these signatures is an important step in confirming that our models for 
planet formation and disk evolution are right and would provide a unique possibility to 
study a newly formed exoplanet and its interaction with the circumstellar material. 

However, the investigations of protoplanetary disks require high-contrast, high-spatial-resolution 
imaging very close to the bright central star. Although in recent years 
numerous groups studied disks around nearby stars at their planet-forming age with the HST,
instrumental limitations do not allow to probe the disk structure 
at scales similar to our inner Solar System. The technique
of polarimetric differential imaging (PDI) as described by Kuhn et al.~(\cite{kuh01})
 is capable of tracing scattered (i.e. polarized) light from a circumstellar
disk very close to the central star.
This promising method enhances the contrast between the disk and the star and, 
thereby, might be important to detect the detailed inner structure of 
protoplanetary disks.

The basic idea of the PDI is to take the difference of two orthogonally 
polarized, simultaeneously acquired images of the same object in order to remove 
all non-polarized light components. The non-polarized light consists mainly of 
the central star's light and the speckle noise, which is -- close to the star -- 
the dominant noise source 
of ground-based observations at optical and near-infrared wavelengths (Racine et al.~\cite{rac99}).
 After this subtraction only the polarized light, such as scattered light from the disk,
is left over. 

To explore the potential of this technique for planet formation studies as well as to demonstrate
the NACO/VLT platform capabilities we targeted
the 8 Myr old (Webb et al.~\cite{web99})  and very close ($56 \pm 7$ pc, Wichmann et al.~\cite{wic98})
classical T Tauri star TW Hya (RA: $11^{\rm h} 01^{\rm m} 51.9^{\rm s}$  Dec:$-34\degr 42' 17''$ J2000). 
At the visual wavelength regime TW Hya shows a polarization variability between 0-3\% (Mekkaden \cite{mek98}).
Being an excellent target for planet formation studies, during the past 
few years several attempts have been made to image its disk as close as possible 
to the star and to identify possible disk structures (Krist et al.~\cite{kri00}, Weinberger et al.~\cite{wei02}).
 In addition,
the presence of a giant planet was predicted (Calvet et al. \cite{cal02}) from modeling the spectral energy distribution 
(SED) (see also Steinacker \& Henning~\cite{ste03}).
 The first application of the PDI to TW Hya 
by Potter (2003\cite{pot03a}, 2003\cite{pot03b}) made use of the
36 element curvature sensing Adaptive Optics (AO) system Hokupa'a (Graves et al.~\cite{gra00}),
of the Gemini North telescope. This H-band data set probed the circumstellar material closer
than 0.5\arcsec to the star.

In the following we describe the first 
results obtained with the PDI technique and the 188 element Shack-Hartmann AO system NACO, 
which is attached to the Nasmyth B
port of the VLT UT4.

\section{Observations and Data Reduction}

The observations were carried out in April 2002 during the commissioning of the NACO/VLT system  
(Lenzen et al.~\cite{len98}, Rousset et al.~\cite{rou98}, Hartung et al.~\cite{har00}) in the Ks band.  In order to simultaneously
 measure two orthogonal components of the polarized light, 
a Wollaston prism was introduced in the light path. The Wollaston prism splits the light into an
ordinary and an extraordinary beam ({\it o}- and {\it e}-beam), separated by 3.5\arcsec{} in the Ks-band.
To eliminate the instrumental polarization and to increase the signal-to-noise ratio (SNR)  of the data, 
a redundant data set with Wollaston angles of 0$\degr$, 45$\degr$, 90$\degr$ and 135$\degr$ was acquired. 
 Each of these images has a field of view (FOV) of 3$\arcsec \times 29.4 \arcsec$ sampled with a pixel size 
 of 0.027$\arcsec$/pixel.
At every Wollaston position a 3-point dithering 
was applied with steps of about 9$\arcsec$ along the field of view to allow sky subtraction and to reduce the influence
of bad pixels. No additional off-source sky frames were
taken.

To ensure the high dynamic range needed for disk analysis, we repeated the complete set of these observations
with defined integration times of 0.4 s and 30 s. The total time spent on the source
 was 24 s and 1800 s in the
short and long exposure series, respectively. As the TW Hya saturated the detector, the AO-performance can not
be accurately measured. However, based on simulations with the NAOS Prepartion Software v1.62 we estimate 
a typical Strehl-ratio of 50\% for these observations.

The data reduction was carried out using self--developed IDL routines. First, every frame was manually 
inspected and those showing 
reflections or electronic ghosts were excluded from  further reduction. 
Hot pixels were efficiently removed by a 3.5-sigma filtering process.
The raw images were sky subtracted and flat field corrected. 
The sky frame was calculated individually for each group of images taken
with the same polarization angle. Following these basic corrections, the ordinary and
extraordinary peaks of the individual exposures were centered and extracted. 
To remove all of the non-polarized intensity (mainly from the central star)
we subtracted images of ordinary polarization from those of extraordinary polarization.

The fine alignment of the individual images before subtraction and co-addition was
carried out by a two-level gaussian fitting procedure (see also Section \ref{Tests}). 

The subtraction process gives the orthogonal $Q_i$ and $U_i$ Stokes components, $i$ standing
for the number of dithering positions. For the 4 angles of the Wollaston prism we
thus derived the values for $Q_i^{0}$, $U_i^{45}$, $Q_i^{90}$ and $U_i^{135}$ (see, e.g. Huard~\cite{hua97}). 
 Using the redundancy of our data set we derived the $Q$ and $U$ mean polarization 
 vectors: \\ 
 $$ Q = {\sum_{i=1}^3 {(Q_i^0 -Q_i^{90}) \over 2}} \, \, {\rm ~and~ }\, \,  U = {\sum_{i=1}^3 {(U_i^{45} -U_i^{135}) \over 2} } $$
  and the polarized intensity:\\ 
  $$ PI= \sqrt{{Q^2 + U^2 \over 2} } .$$

This reduction procedure was repeated independently on both the long and 
the short exposure series, resulting
in the  $PI_{\rm Short}$ and $PI_{\rm Long}$ polarized intensity maps.

 Due to the brightness of TW Hya, the inner regions of the raw frames
 exceeded the regime of linear detector response. These data points
 ($r < 0.06\arcsec$ in the short exposure series and $r<0.45\arcsec$ in the long exposure series)
 were excluded from the data reduction and further analysis.
 The field of view (FOV) was limited to $\sim3\arcsec$ by the field mask. 
  During the dithering process slight movements perpendicular to the dithering directions
  occured. Since the resulting FOV is the intersection of the FOVs of the individual frames, 
the final field of view was reduced to $\sim 0.54\arcsec$ and $ \sim 1.43\arcsec$ for the 
 short and long exposure series, respectively.
 
 To characterize the surface brightness distribution of the disk we measured the radial profile
 of the polarized intensity, derived from 3-pixel-wide apertures, roughly equivalent to the full width at half
 maximum (FWHM) of the point spread function (PSF).

\begin{figure*}
   \centering
   \includegraphics[width=8cm]{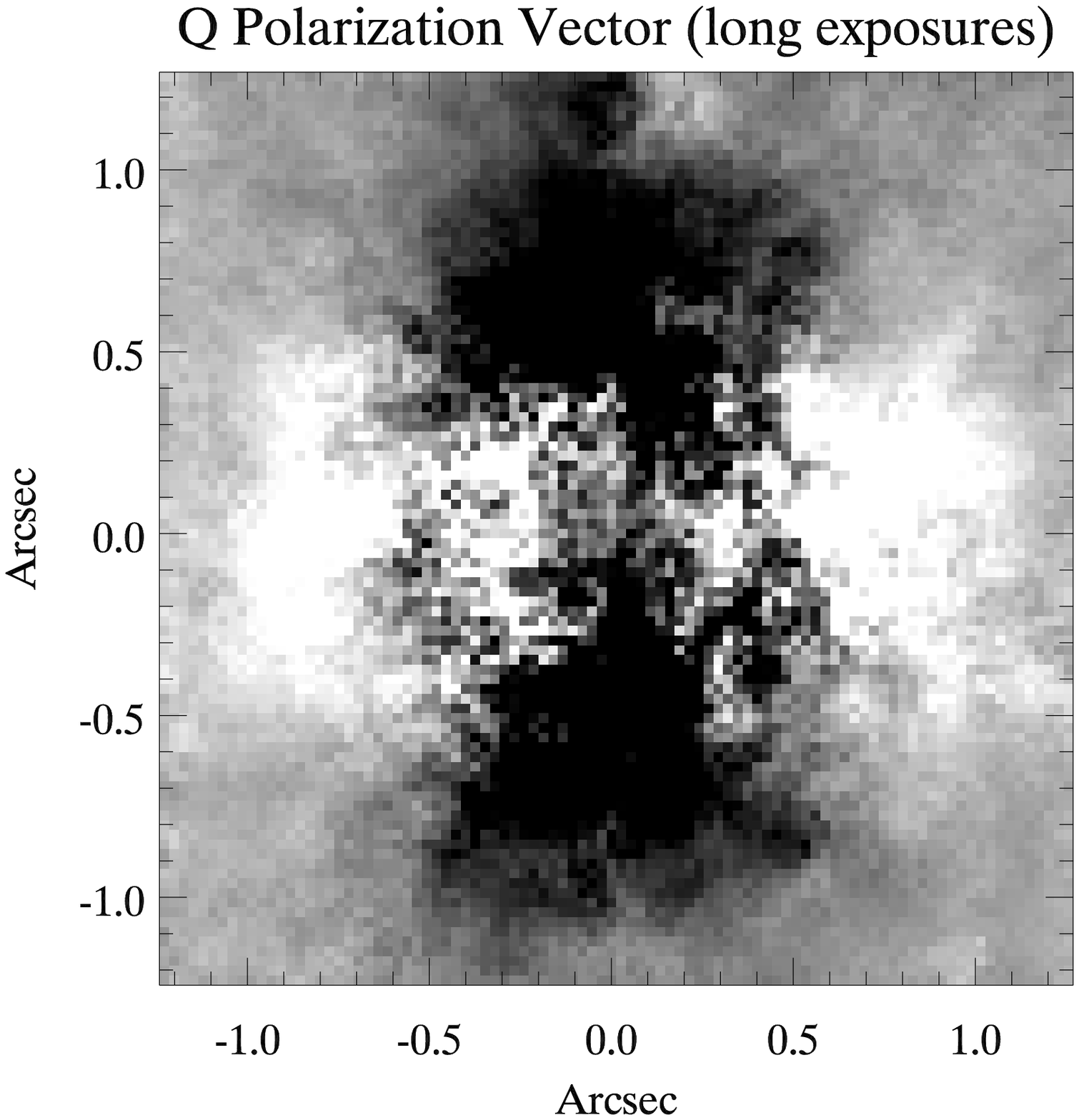}
   \includegraphics[width=8cm,height=7.2cm]{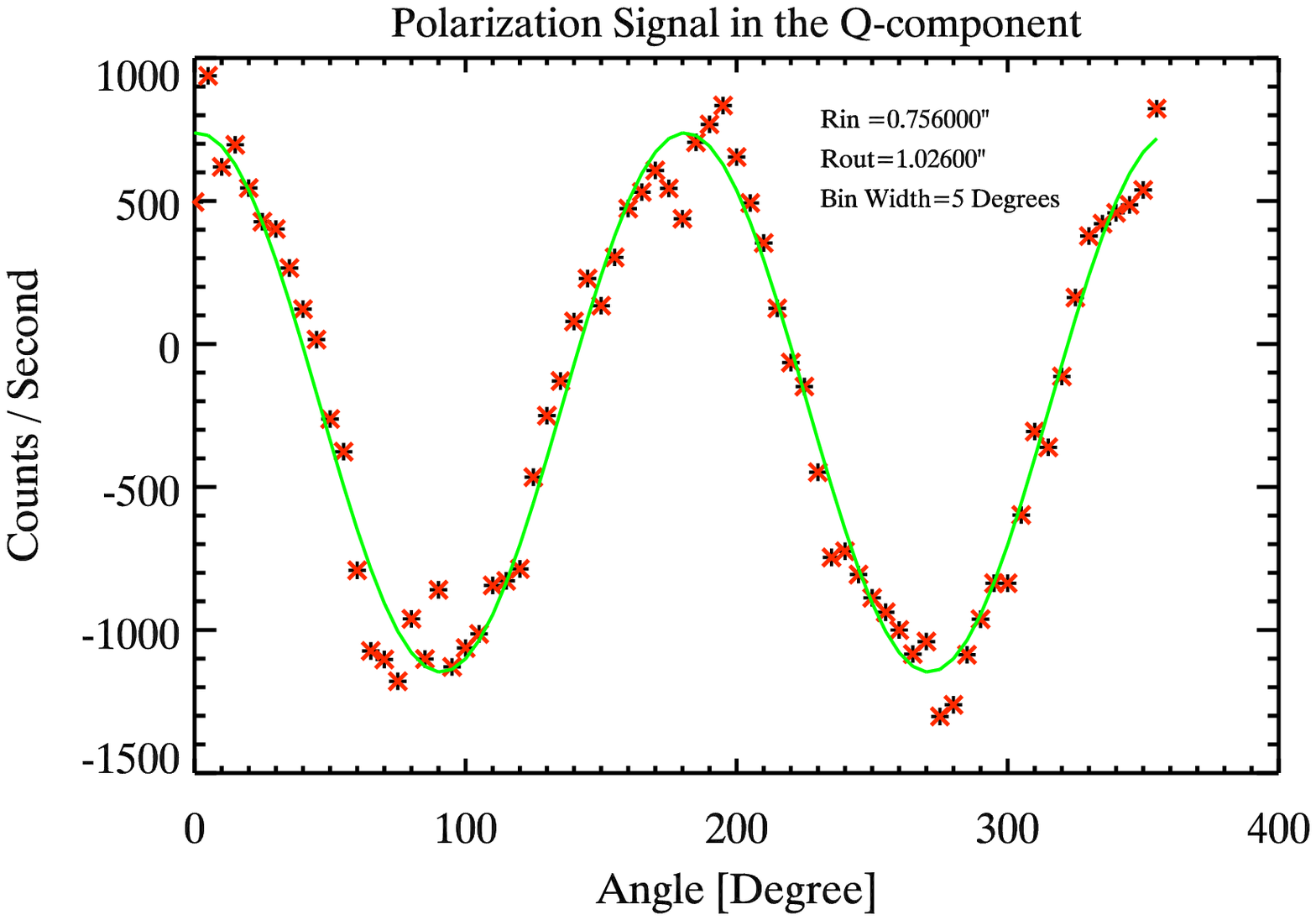}
    \caption{{\it Left panel:}{\rm The characteristic {\it butterfly}} pattern of an 
    extended polarized source around
TW Hya in the $Q_{\rm Long}$ image.  This pattern
can be identified between 0.5\arcsec{} and 1.4\arcsec{} from the star on both the Q- and U (long exposure) images.
The image is centered on TW Hya.
{\it Right panel:} Counts in the Q-component of the polarization vector as a function
of position angle in an annulus between 0.75\arcsec{} and 1\arcsec{} from the 
long exposure series. The asterisks
mark the data points averaged over 5 degree bins, while the curve is the
best fitting frequency-fixed cosine. 
 The strong sinusoidal modulation indicates that the light-scattering 
 dust is distributed nearly axisymmetrically around the light source.
  }
\label{lepkelong}
\end{figure*}  

\begin{figure*}
   \centering
   \includegraphics[width=8cm]{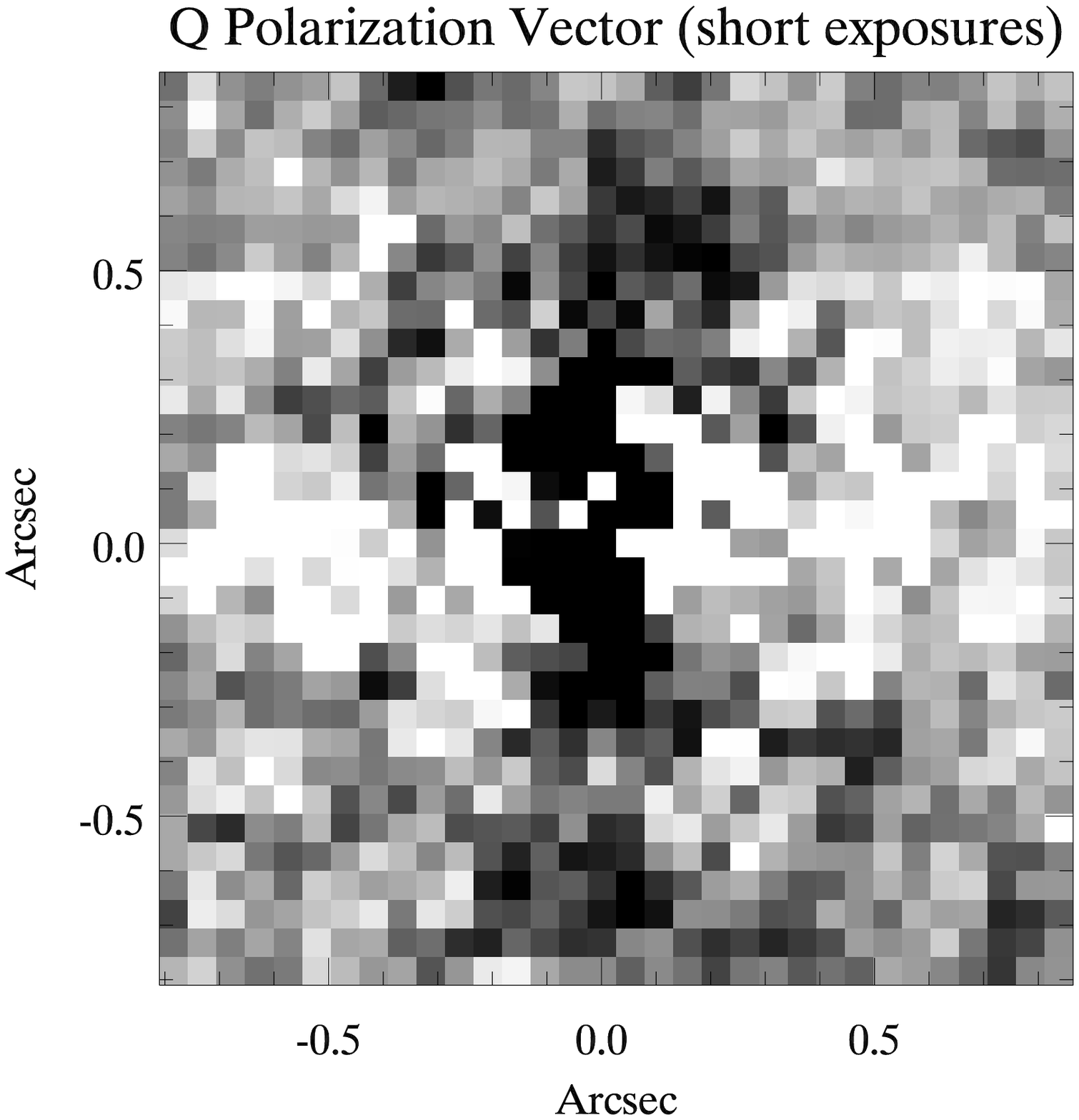}
   \includegraphics[width=8cm,height=7.2cm]{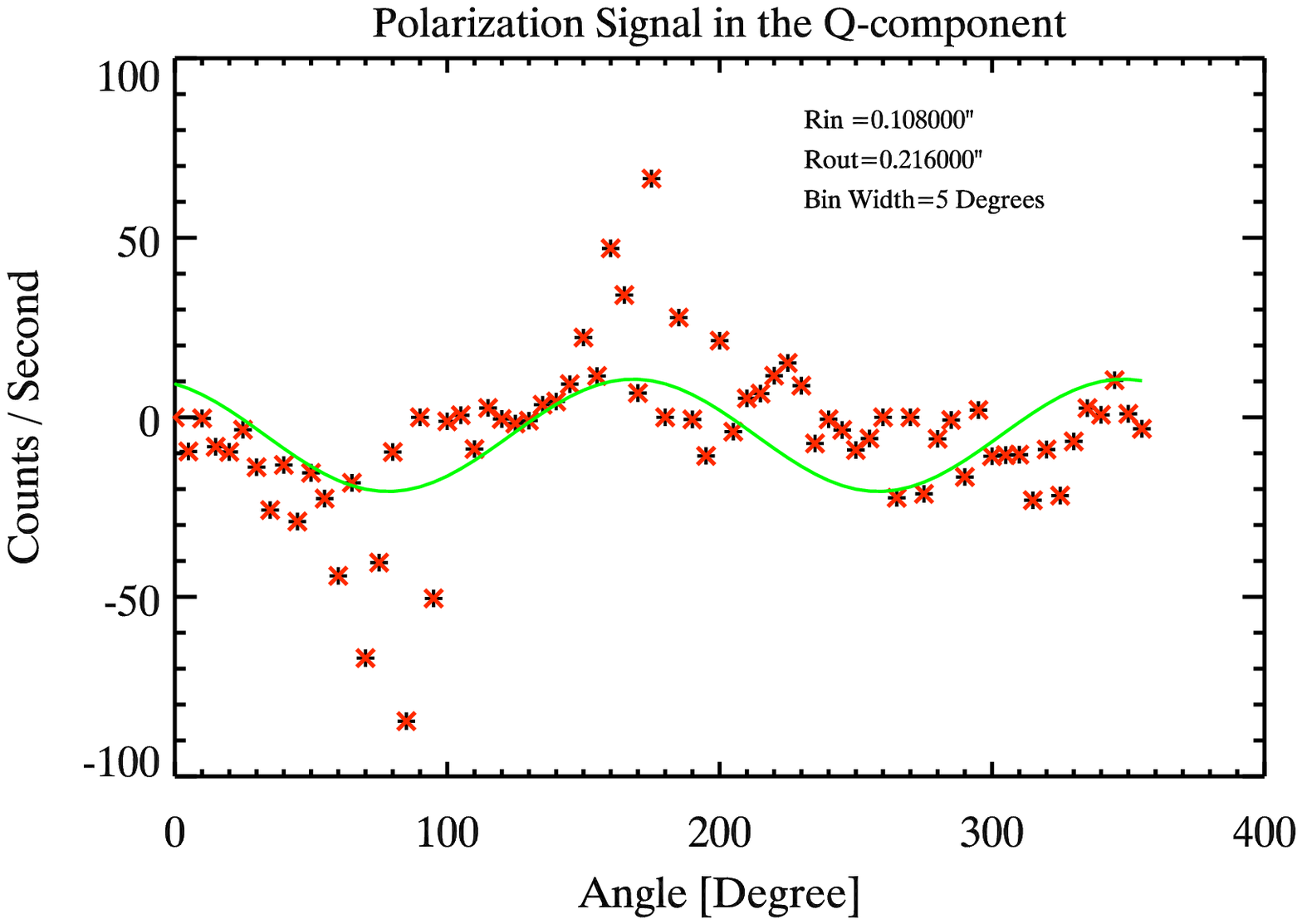}
    \caption{{\it Left panel:} Same as left panel of Fig. 1 but for the 
 the $Q_{\rm Short}$ image.  The polarization pattern
can be identified between 0.1\arcsec{} and 0.4\arcsec{} on the short exposure image. 
{\it Right panel:} Same as Fig. \ref{lepkelong} for an annulus between 
0.1\arcsec{} and 0.2\arcsec{} from the short exposure series.
 The sinusoidal modulation (with correct phase and frequency) shows the presence
 of light-scattering  dust as close as 0.1\arcsec{} to the star.
 For a comparison of the residuals from  the PSF-comparison star plotted in the 
 same scale, see right panel of Fig.~\ref{UandStd}.}
\label{lepke}
\end{figure*}  

\begin{figure*}
   \centering
   \includegraphics[width=8cm, height=7.2cm]{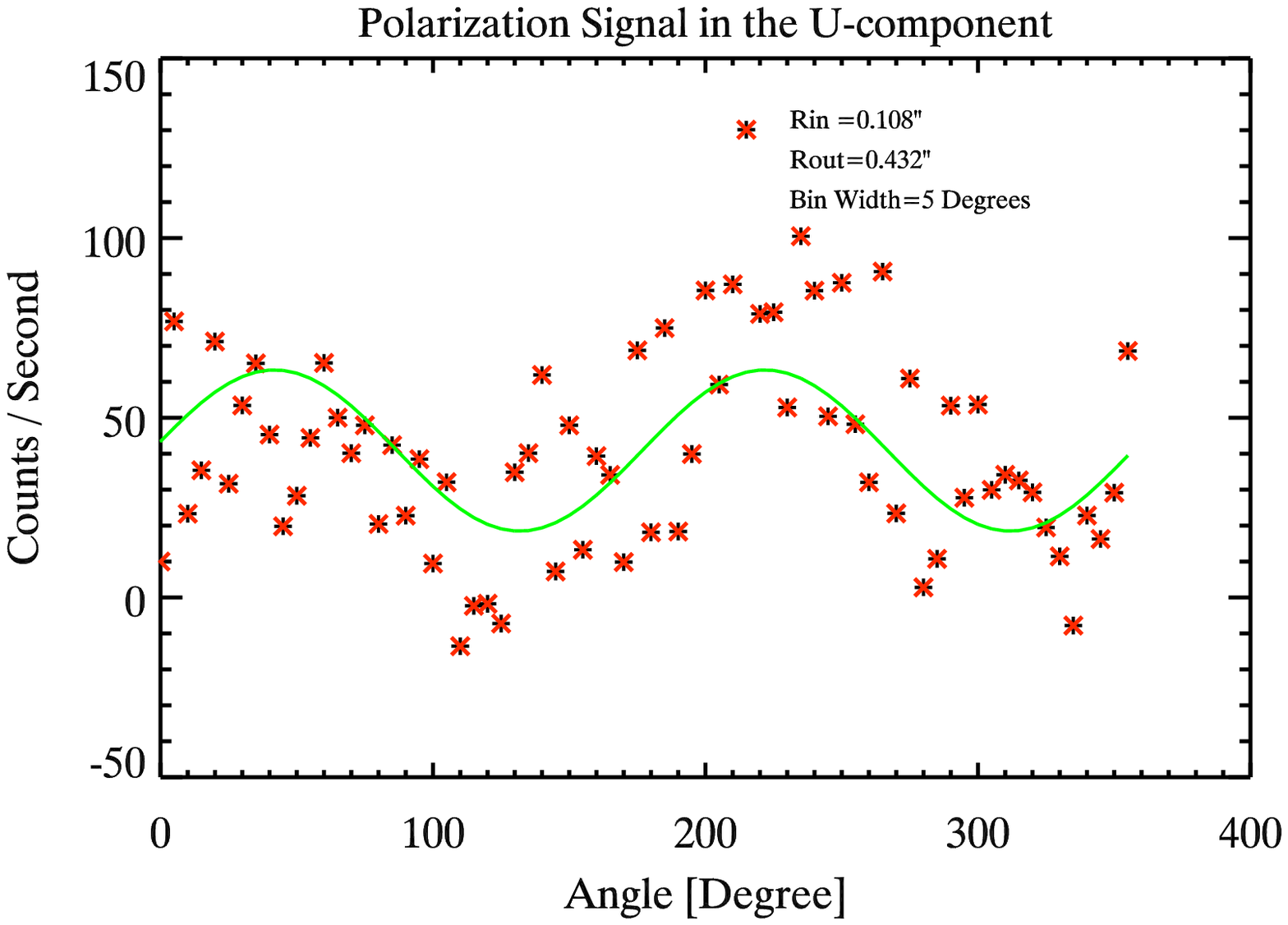}
   \includegraphics[width=8cm,height=7.2cm]{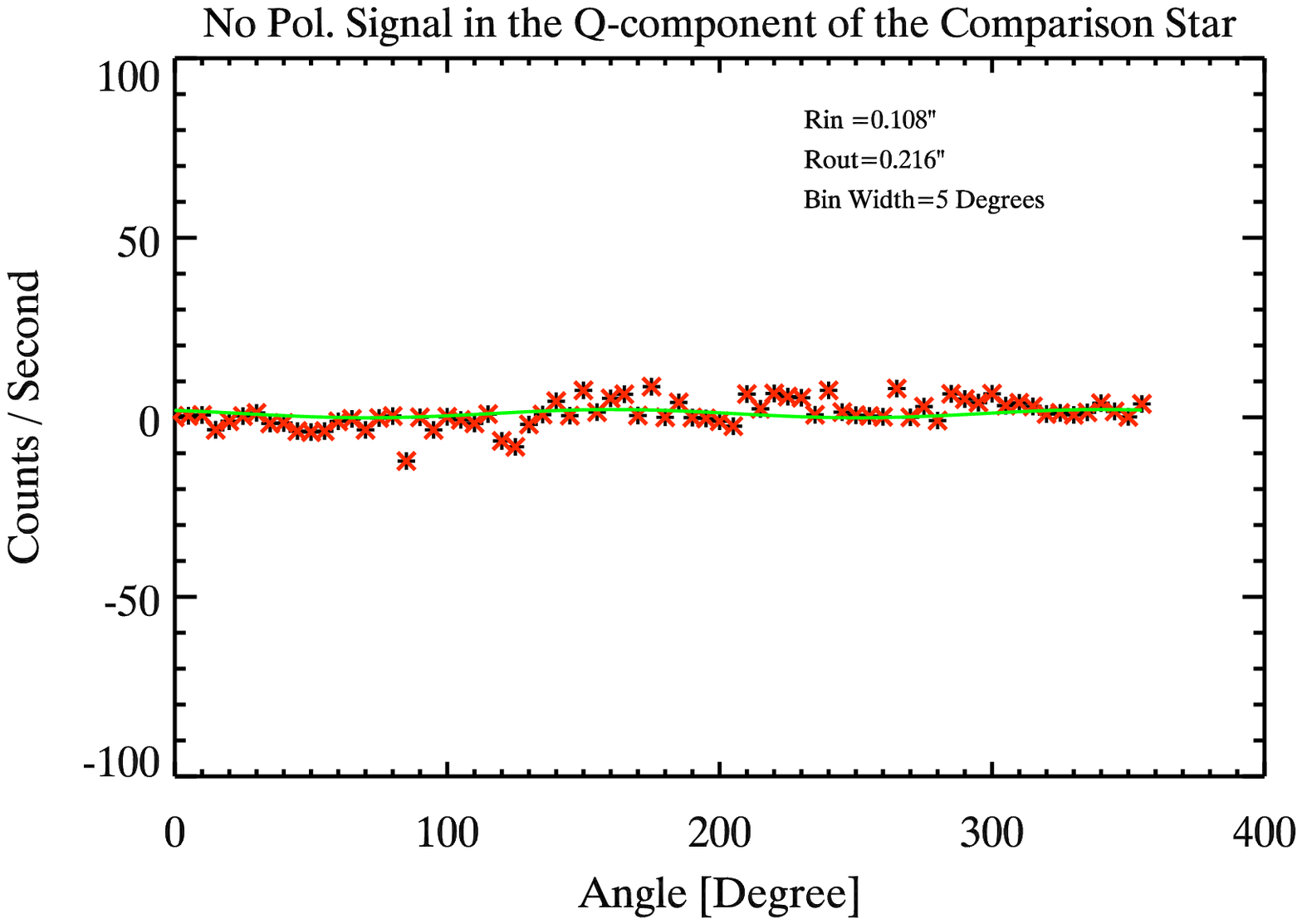}
    \caption{ {\it Left panel:} TW Hya Short U image -- pol. signal between 
    0.1\arcsec{} and 0.4\arcsec{}. 
    {\it Right panel:} Similar plot for the PSF comparison star 
    GSC 07208-00319. No polarization signal visible (the plot range and
    analysis parameters are set to be the same as for the right panel 
    in Fig.~\ref{lepke}).}
\label{UandStd}
\end{figure*}

\begin{figure*}
   \centering
   \includegraphics{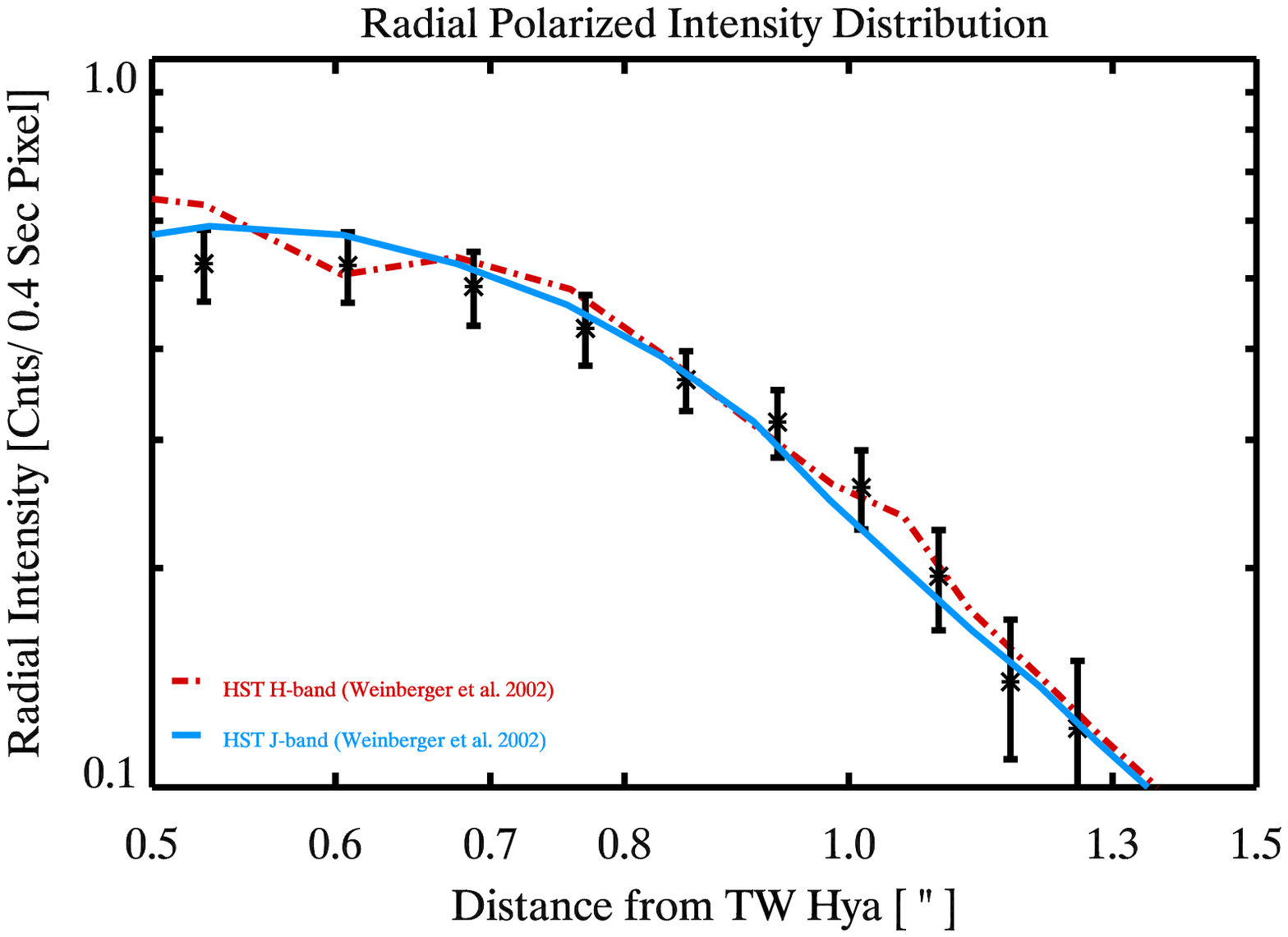}
\caption{Relative radial polarized intensity distribution of the TW Hya disk between
0.5\arcsec{} and 1.4\arcsec{}. The asterisks mark our 3-pixel-binned Ks-band PDI data, while
the  solid and dashed curves show the J- and H-band surface brightness 
distributions from Weinberger et al. (\cite{wei02}). Although the surface brightness is not equivalent
to the PI, applying an arbitrary offset shows the similar behaviour of the three curves.
 The slope change around 0.8\arcsec{} has been 
found already by Krist et al. (\cite{kri00}) and Weinberger et al. (\cite{wei02}). 
The error bars represent the combined errors from the PSF-comparison 
residuals and the statistical errors of each annulus.}
\label{profile}%
\end{figure*}

\section{Signal Analysis}

 For detecting the polarization
 signal as close as possible to the star a very careful data inspection is essential -- 
 residuals from the subtractions as well as sub-pixel misalignments between the
 individual frames are the dominant noise sources. 
 While the scalar PI image contains no information
 on the direction of the polarization, the Q- and U-components are projections
 of the polarization vector and thus show a sinusoidal modulation with the
 angle in respect to the polarization direction of the Wollaston-prism (i.e. 
 a maximum at the angles
 parallel to the polarization direction of the Wollaston prism and a minimum perpendicular
 to it). In the case of axisymmetrically distributed dust this modulation leads to 
 the {\it butterfly} pattern as shown in the right panel of  Fig. \ref{lepkelong} and described in Kuhn et al. (\cite{kuh01}).
 By using the information about the polarization orientation and by comparing this pattern to what is expected
 from an axi-symmetric light scatterer, we can better discriminate between noise and signal.

  The right panel of Fig. \ref{lepkelong} shows the Q-component of the polarization signal radially 
  integrated (over 5-degree bins) in an annulus between 0.75\arcsec{} and 1.0\arcsec{} 
  from the long exposure series.  
  A very similar, but $\pi$/2-shifted modulation is evident
  in the U-images.  The clearly 
 sinusoidal modulation arises from the strong polarization signal. When probing the
 extent of the  polarized circumstellar material, we looked for the presence of a 
 similar pattern (phase- and frequency-correct sinusoidal modulation) as a function
 of radial distance from TW Hya.
 
 We found an extended polarization pattern with a high signal-to-noise ratio between
 0.5 and 1.4\arcsec{} distance from the star in both the Q and U long exposure images.  In the short exposure
 images we identified 
 light-scattering material between 0.2\arcsec{} and 0.4\arcsec{} from
 TW Hya in both the Q- and U-components. 
 Inspection of the higher signal-to-noise Q-component (short exposure) showed this pattern 
 also between 0.1\arcsec and 0.2\arcsec{} (see the left panel of Fig. \ref{lepke}). 
 
 In order to estimate the contrast ratio between the polarized and the unpolarized fluxes
 we compared the fluxes of the direct images (after sky subtraction and bad pixel removal) to the
 amplitude of the fitted cosine. The typical contrast ratios are $F_{P/U}\simeq$1\% and $F_{P/U}\simeq$3.5--4\%
 for radii of $r = 0.1\arcsec$ --~$0.3\arcsec$ and $r\simeq1\arcsec$, respectively.

\section{Reliability Tests and Confirmation}    
  \label{Tests}
 
  To confirm our results, we analyzed the Q- and U-components
  of the PDI data set of the PSF-comparison star GSC 07208-00319, which was observed
  and reduced in an identical way to the TW Hya data set. In this case neither polarized 
  emission in the PI image nor any sign of sinusoidal modulation in the Q- and U-band
  images could be found (see right panel of Fig.~\ref{UandStd}).
   
 We also conducted extensive tests to check the reliablity of our data reduction and analysis toolkit.
 During these test runs the effects of the following errors were investigated: 
 sub-pixel shifts between the individual frames, different 
 centering, interpolation, and flat field normalization methods. For each test run the
 complete data reduction and analysis has been executed both for TW Hya and the PSF 
 comparison star.
 
 As the possible sub-pixel misalignments might produce a bipolar, positive-negative pattern 
 mimicking the butterfly pattern of the extended polarization signal, special care was taken
 to investigate their effects on our results. In order to simulate centering problems, 
 we have introduced normally distributed random shifts with FWHMs
  of 0.2, 0.5, 1.0 and 1.5 pixels {\it after} the centering step. The sinusoidal 
  modulation of the TW Hya short exposure frames gets strongly distorted and asymmetric 
  even for the sub-pixel misalignments (0.2 and 0.5) and the pattern is 
  completely disrupted for misalignments as large as 1 or 1.5 pixels. The short exposure 
  images of the PSF comparison star start to show a sinusoidal modulation, but its 
  frequency is {\it one half} of that of the polarization signal. These images show, 
  in fact, offset negative and positive peaks which can  easily be discriminated from the
  butterfly pattern shown in Fig.~\ref{lepkelong}. The long exposure images are less sensitive
  to sub-pixel misalignment but are noticeably distorted in the case of large (1 or 1.5 pixel)
  centering errors.
  
  In order to explore the influence of systematic errors in the centering procedure
  we have introduced fixed shifts of 0.1 and 0.2 pixels after our centering procedure.
  We found that systematic centering errors as small as 0.1 pixels lead to a strong
  positive-negative pattern and the disruption of the polarization signal. These systematic
  errors produced similar pattern for the PSF comparison star as reported for the case of
  random shifts and can, therefore, not resemble the correct polarization pattern.
  
  Additional tests included the comparison of centering methods based on 
  cross-correlation and on Gauss fitting, bilinear and cubic interpolations at 
  the rotation of the images, as well as the local and global normalization of the flat field images. 
  We found that the Gauss-fitting centering, the bilinear interpolation and the individual 
  normalization of each stripe in the Wollaston-masked flat field leads to the highest 
  signal-to-noise ratios.
   
 These test runs proved that the strongest noise sources in the reduction procedure
 are the subpixel misalignments and the  flat field errors. We found no combination 
 of the artificially introduced errors that could create a false  
 (phase- and frequency-correct) sinusoidal modulation and thus could lead to 
 the misinterpretation of our data set.
 
 Chromatic effects from the Wollaston prism could also lead to slight differences
 between the o- and e-beams and thus to an imperfect subtraction. However, the 
 MgF$_{2}$ prism used in the CONICA camera introduces a 
 relative displacement of only 86 mas between the 
 blue and red edge of the broad K-band (CONICA Manual). In the narrower Ks-band 
 this effect is even smaller and is therefore not expected to influence the  
 PDI observations.

 As TW Hya and the PSF star were observed in very similar airmass ranges 
 (1.02 -- 1.08 and 1.09 -- 1.22) and seeing conditions (0.6\arcsec), we expect that any
 atmospheric effect leading to a false polarization signal would effect the PSF 
 comparison images in the same way. Since there is no evidence for such an atmospheric 
 influence in the comparison images (see right panel of Fig.~\ref{UandStd}), 
 we exclude atmospheric chromatism as a possbility
 to mimick PDI signals.
 
 The examination of the data reduction, possible instrumental and atmospheric
 effects led us to conclude that the extended polarization signal in our images indeed
 originate from the light scattering on the circumstellar dust around TW Hya.

\section{Results} 

Based on the NACO observations, we obtained Ks-band images of the circumstellar disk around
the classical T Tauri star TW Hya, which probe the disk structure closer to the star than
any previous observation. Our diffraction limited PDI of TW Hya
shows an extended butterfly pattern (see Figs. \ref{lepkelong}, \ref{lepke}) characteristic for a spatially
resolved axisymmetric light scattering source. This pattern is present between 
0.5\arcsec{} and 1.4\arcsec{} on the long exposure Q- and U-images and between 0.1\arcsec{} and 0.4\arcsec{}
in the short exposure images. 
  The observations of the regions closer than 0.1\arcsec{} to the star and between 0.4\arcsec{} and 0.5\arcsec{} do not 
 have sufficient SNR for a reliable analysis.
The SNR of the data set is sufficiently high to plot the azimuthally 
integrated Ks-band surface brightness density distribution between 0.5\arcsec{} 
and 1.4\arcsec{} (see Fig.~\ref{profile}).

\section{Discussion}

\subsection{Scattering}

The presence of the extended polarization signal across our FOV confirms the existence
of  light scattering material. The sinusoidal modulation of the Q- and U-images 
confirms the centrosymmetric polarization pattern.
The fact that we detected such emission even as close as
$\sim$0.1\arcsec{} ($\sim$6 AU) to TW Hya is the first direct evidence that its dust disk extends 
so far in. This detection does not support the model of
Krist et al. (\cite{kri00}) who proposed the presence of an inner 'dark' zone 
($r\simeq 18$ AU = 0.3\arcsec{}) 
with suppressed flaring angle, necessary to explain the exceptional brightness of the TW Hya disk.  

 Based on the modeling of the SED, Calvet et al. (\cite{cal02}) predicted
the presence of a mostly dust-evacuated inner gap at $r<3-4$  AU ($r<0.07''$). Such a gap could be a 
tell-tale signpost of a massive planet in close orbit around TW Hya and would be of high
importance. However, our commissioning data do not allow to draw any definite conclusions on the 
existence of this feature.

\subsection{Radial Profile and Slope Change}

In Fig.~\ref{profile} we show the comparison of the radial polarized intensity derived from our PDI observation and
(arbitrarily shifted) radial surface brightness profiles measured by the HST (Weinberger \cite{wei02}).  
Although these measurements
probe different quantities at different wavelengths (Ks and J, H,  respectively)  a remarkable similarity is
evident between the radial slopes. Based on this similarity, we conclude that the
polarization degree is nearly independent of the radius and thus the polarized intensity characterizes the
surface brightness. Thus, the PDI method can lead to contrast enhancement without the loss of information.

 The radial surface brightness profile between $0.9\arcsec{}$ and $1.4\arcsec{}$ can be well fitted by a
power-law function $I_{\rm Ks,1} \propto r^{-3.1 \pm 0.3}$. This is slightly steeper than the 
$I_{\rm J,H} \propto r^{-2.6 \pm 0.1}$ 
behaviour found by Weinberger et al. (\cite{wei02}) at 1.1 and 1.6 $\mu$m, 
for the region between 0.8\arcsec{} and 2.7\arcsec{}.

To place our results in context, we compare them to surface brightness slopes
predicted by simple analytical approximations (e.g. Whitney \& Hartmann \cite{whi92}).
These approximations consider only isotropic scattering and
simple geometric effects and do not include wavelength-dependent terms. 
The predicted surface brightness power-law exponent for an optically thick, geometrically thin disk
is $-3$, while for a flaring disk an exponent of -2 is expected.
 Our Ks-band results in the $0.9\arcsec{} < r < 1.4\arcsec{}$ 
region support the flat disk geometry or a very small flaring. 
However, the surface brightness slope is determined
by an intricate interplay between many disk and dust parameters and is not a conclusive measure 
of the disk geometry.
 
 As to underline this statement, a strong, gradual change in the brightness profile around 0.8\arcsec{} from TW Hya is
evident. The annulus between $0.5\arcsec{} < r < 0.7\arcsec{}$ is characterized by the very gentle 
surface brightness behaviour of $I_{\rm Ks,2} \sim r^{-0.9\pm0.2}$. The slope of the curve 
is even less steep than that expected from a flared disk (Whitney \& Hartmann~\cite{whi92}).

Although the radial brightness slope is strongly influenced by the disk geometry,
slope changes can have different origins. Even though a detailed analysis of 
the reasons of the slope changes is beyond the scope of the current paper, we refer 
to some of them. The most obvious one is an actual change in the surface density slope of the
disk, translating into a slope change in the surface brightness. Such clear-cut changes 
might be the results of external or internal perturbation of the disk by a companion or 
a massive planet as seen in the case of HR 4796A (see, e.g. Boccaletti et al.~\cite{boc03}).

 A second cause might be the change of the flaring angle and through this the change in the 
illumination of the disk's surface. Alternatively, a radial dependence of the scattering cross
section of the dust grains might introduce modulations in the surface brightness profile.
Disentangling the effects of these processes requires high-angular resolution 
observations at longer wavelengths.

We note, that this change in the slope at 0.8\arcsec{} has been detected previously 
at shorter wavelengths by HST 
direct imaging and coronographic studies (Krist et al.~\cite{kri00}, Weinberger et al.~\cite{wei02}), 
together with other slope changes in the outer parts ($>$1.4\arcsec) of the disk.  

Our data were acquired during the commissioning of the NACO instrument and
suffer from high detector noise and limited AO performance. The
current performance of NACO is expected to result in roughly five times better
SNR. 

The first results of a PDI survey of the TW Hya Association  are
presented by Hu\'elamo \& Brandner (\cite{hue03}).

For future PDI observers it might be important 
to point out some of the limiting factors in our data set. 
First, the dynamic range of the data set has to be increased by applying
series of different exposure times. Furthermore, the flat field 
calibration plays a critical role and should be carried out with great care.
Finally, the neccessarily saturated star will pose limitations to 
the accurate alignment of the individual frames, which will in practice 
determine the inner radius until which the polarization signal can be detected.

\section{Summary}

 We presented the first high-contrast observations carried out with NACO/VLT using
the PDI technique. The extended scattering pattern of TW Hya provides the first direct 
proof that the disk extends closer than 0.5\arcsec{}, up to $\sim$ 0.1\arcsec{} ($\sim$ 6 AU) 
from the star.

 We derive the first Ks-band radial polarized intensity profile between
0.5\arcsec{}  and 1.4\arcsec{}. This  profile strongly resembles the surface brightness profile
seen at shorter wavelengths and shows as well a strong, gradual slope change
around 0.8\arcsec{}. The polarized intensity profile
between 0.9\arcsec{} and 1.4\arcsec{} stands close to that expected from a flat disk.

 These results demonstrate the potential of the PDI technique at the NACO/VLT 
 platform for future planet formation studies. This technique is capable of
 imaging a new, beforehand unaccessible regime of  circumstellar disks 
 at very high resolution. Such images can help understanding the evolution
 of protoplanetary disks at the scales comparable to our inner Solar System. 
 
\begin{acknowledgements} 
 We acknowledge the outstanding support of the ESO staff
both in Garching and at Paranal to the construction and
commissioning of NACO. We thank O. Sch\"utz for his useful comments 
on the manuscript. The suggestions of 
the referees (R. Racine and P. Bastien) helped to improve the clarity of the paper.
\end{acknowledgements}

\end{document}